**Title:** A Causal Roadmap for Generating High-Quality Real-World Evidence

**Authors and Affiliations:** Dang LE[1], Gruber S[2], Lee H[3], Dahabreh I[4], Stuart EA[5], Williamson BD[6], Wyss R[7], Díaz I[8], Ghosh D[9], Kıcıman E[10], Alemayehu D[11], Hoffman KL[12], Vossen CY[13], Huml RA[14], Ravn H[15], Kvist K[15], Pratley R[16], Shih MC[17,18], Pennello G[19], Martin D[20], Waddy SP[21], Barr CE[22,23], Akacha M[24], Buse JB[25], van der Laan M*[1], Petersen M*[1]
**\*Co-Senior Authors**

[1]Department of Biostatistics, University of California, Berkeley, CA, USA

[2]TL Revolution, Cambridge, MA, USA

[3]Office of Biostatistics, Office of Translational Sciences, Center for Drug Evaluation and Research, U.S. Food and Drug Administration, Silver Spring, MD, USA

[4]Department of Epidemiology, Harvard T.H. Chan School of Public Health, Boston, MA, USA

[5]Department of Mental Health, Johns Hopkins Bloomberg School of Public Health, Baltimore, MD, USA

[6]Biostatistics Division, Kaiser Permanente Washington Health Research Institute, Seattle, WA, USA

[7]Division of Pharmacoepidemiology and Pharmacoeconomics, Brigham and Women's Hospital, Harvard Medical School, Boston, MA, USA

[8]Division of Biostatistics, Department of Population Health, New York University Grossman School of Medicine, New York, NY, USA

[9]Department of Biostatistics and Informatics, Colorado School of Public Health, University of Colorado Anschutz Medical Campus, Aurora, CO, USA

[10]Microsoft Research, Redmond, WA, USA

[11]Global Biometrics and Data Management, Pfizer Inc., New York, NY, USA

[12]Department of Epidemiology, Mailman School of Public Health, Columbia University, New York, NY, USA

[13]Syneos Health Clinical Solutions, Amsterdam, The Netherlands

[14]Syneos Health Clinical Solutions, Morrisville, NC, USA

[15]Novo Nordisk, Søborg, Denmark

[16]AdventHealth Translational Research Institute, Orlando, FL, USA

[17]Cooperative Studies Program Coordinating Center, VA Palo Alto Health Care System, Palo Alto, CA, USA

[18]Department of Biomedical Data Science, Stanford University, Stanford, CA, USA

[19]Division of Imaging Diagnostics and Software Reliability, Office of Science and Engineering Laboratories, Center for Devices and Radiological Health, U.S. Food and Drug Administration, Silver Spring, MD, USA

[20]Global Real World Evidence Group, Moderna, Cambridge, MA, USA

[21]National Center for Advancing Translational Sciences, Bethesda, MD, USA

[22]Graticule Inc., Newton, MA, USA

[23]Adaptic Health Inc., Palo Alto, CA, USA

[24]Novartis Pharma AG, Basel, Switzerland

[25]Division of Endocrinology, Department of Medicine, University of North Carolina, Chapel Hill, NC, USA

**Corresponding Author:**
Lauren Eyler Dang
Division of Biostatistics
University of California, Berkeley
2121 Berkeley Way, Room 5302
Berkeley, CA 94720
(510) 642-3241
Lauren.eyler@berkeley.edu

**Conflict of Interest Statement:**

**LED** reports tuition and stipend support from a philanthropic gift from the Novo Nordisk corporation to the University of California, Berkeley to support the Joint Initiative for Causal Inference. **EK** is employed by Microsoft. **DA** is employed by Pfizer Inc. and holds stocks in Pfizer Inc. **CYV** and **RAH** are employed by Syneos Health. **CYV** reports that her husband is employed by Galapagos. **HR** and **KK** are employed by Novo Nordisk A/S and own stocks in Novo Nordisk A/S. **RP** has received the following (directed to his institution): speaker fees from Merck and Novo Nordisk; consulting fees from Bayer AG, Corcept Therapeutics Incorporated, Dexcom, Endogenex, Inc., Gasherbrum Bio, Inc., Hanmi Pharmaceutical Co., Hengrui (USA) Ltd., Lilly, Merck, Novo Nordisk, Pfizer, Rivus Pharmaceuticals Inc., Sanofi, Scohia Pharma Inc., and Sun Pharmaceutical Industries; and grants from Hanmi Pharmaceuticals Co., Metavention, Novo Nordisk, and Poxel SA. **DM** is employed by Moderna. **CEB** is co-founder of Adaptic Health Inc., Managing Director of Pivotal Strategic Consulting, LLC, and receives consulting fees from Graticule Inc. and Sophic Alliance Inc. **MA** is employed by Novartis Pharma AG. **JBB** reports contracted fees and travel support for contracted activities for consulting work paid to the University of North Carolina by Novo Nordisk; grant support by Dexcom, NovaTarg, Novo Nordisk, Sanofi, Tolerion and vTv Therapeutics; personal compensation for consultation from Alkahest, Altimmune, Anji, AstraZeneca, Bayer, Biomea Fusion Inc, Boehringer-Ingelheim, CeQur, Cirius Therapeutics Inc, Corcept Therapeutics, Eli Lilly, Fortress Biotech, GentiBio, Glycadia, Glyscend, Janssen, MannKind, Mellitus Health, Moderna, Pendulum Therapeutics, Praetego, Sanofi, Stability Health, Terns Inc, Valo and Zealand Pharma; stock/options in Glyscend, Mellitus Health, Pendulum Therapeutics, PhaseBio, Praetego, and Stability Health; and board membership of the Association of Clinical and Translational Science. **MvdL and SG** report that they are co-founders of the statistical software start-up company TLrevolution, Inc. **MvdL and MP** report personal compensation for consultation from Novo Nordisk.

**Abstract:** Increasing emphasis on the use of real-world evidence (RWE) to support clinical policy and regulatory decision-making has led to a proliferation of guidance, advice, and frameworks from regulatory agencies, academia, professional societies, and industry. A broad spectrum of studies use real-world data (RWD) to produce RWE, ranging from randomized controlled trials with outcomes assessed using RWD to fully observational studies. Yet many RWE study proposals lack sufficient detail to evaluate adequacy, and many analyses of RWD suffer from implausible assumptions, other methodological flaws, or inappropriate interpretations. The *Causal Roadmap* is an explicit, itemized, iterative process that guides investigators to pre-specify analytic study designs; it addresses a wide range of guidance within a single framework. By requiring transparent evaluation of causal assumptions and facilitating objective comparisons of design and analysis choices based on pre-specified criteria, the *Roadmap* can help investigators to evaluate the quality of evidence that a given study is likely to produce, specify a study to generate high-quality RWE, and communicate effectively with regulatory agencies and other stakeholders. This paper aims to disseminate and extend the *Causal Roadmap* framework for use by clinical and translational researchers, with companion papers demonstrating application of the *Causal Roadmap* for specific use cases.

**Introduction**

The 21st century has witnessed a dramatic increase in the quality, diversity, and availability of real-world healthcare data (RWD) in forms such as electronic health records and registry or claims databases[1]. In 2016, as part of a strategy to improve the efficiency of medical product development, the United States Congress passed the 21st Century Cures Act[2] that mandated the development of United States Food and Drug Administration (FDA) guidance on potential regulatory uses of real-world evidence (RWE) – defined as "*clinical* evidence about the usage and potential benefits or risks of a medical product derived from analysis of RWD"[3]. Internationally, stakeholders including other regulatory agencies, industry, payers, academia, and patient groups have also increasingly endorsed the use of RWE to support regulatory decisions[4,5]. Emerging sources of RWE under evaluation include pragmatic clinical trials, externally controlled trials or hybrid randomized-external data studies, and long-term follow-up studies[6–8].

There are multiple motivations for generating RWE. First, RWE has long been used in post-market safety surveillance to uncover the presence of rare adverse events not adequately evaluated by phase III randomized controlled trials (RCTs) for reasons including strict eligibility criteria, strict treatment protocols, limited patient numbers, and limited time on treatment and in follow-up[9]. Second, recent drug development efforts have more commonly targeted rare diseases or conditions without effective treatments[10]. RWD can be useful in such contexts when it is not practical to randomize enough participants to power a standard RCT or when there is an ethical imperative to minimize the number of patients assigned to the trial control arm[11,12]. RWE was also highly valuable during the COVID-19 pandemic; observational studies reported timely evidence of vaccine booster effectiveness[13,14], compared the effectiveness of different vaccines[15], and evaluated vaccine effectiveness during pregnancy[16].

Despite the many ways in which RWE may support policy or regulatory decision-making, the prospect of erroneous conclusions resulting from potentially biased effect estimates has led to appropriate caution when interpreting the results of RWE studies. One concern is data availability; data sources might not include all relevant information for causal estimation even in randomized studies that generate RWE. Another concern is lack of randomized treatment allocation in observational RWE. These issues create challenges for estimating a causal relationship outside of the "traditional" clinical trial space.

In an attempt to guide investigators towards better practices for RWE studies, there has been a blossoming of input from regulatory agencies, academia, and industry in the form of guidelines and frameworks addressing different stages of the process of evidence generation[3,5,17–23]. Yet incoming submissions to regulatory agencies lack standardization and consistent inclusion of all information that is relevant for evaluating the quality of evidence that may be produced by a given RWE study[20]. How, then, can we help investigators do a better job of estimating causal effects – and evaluating the plausibility of assumptions needed to estimate causal effects – based at least partially on RWD?

To help answer this question, the Forum on the Integration of Observational and Randomized Data (FIORD) meeting was held in Washington, D.C. November 17-18, 2022 to discuss perspectives from regulatory and federal medical research agencies, industry, academia, trialists, methodologists, and software developers. FIORD participants discussed their experiences with RWE guidance and best practices and identified necessary steps and priorities for broadening usage by investigators. Specifically, participants determined the need for a unifying structure to assist with specification of a complete analytic design for an RWE study, including both the statistical analysis plan and additional design elements relevant for optimizing and evaluating the quality of evidence produced.

The *Causal Roadmap*[24–30] (hereafter, the *Roadmap*) addresses this need because it is a general, adaptable framework for causal and statistical inference that is applicable to all studies that generate RWE, including studies with randomized treatment allocation and prospective and retrospective observational designs. It is consistent with existing guidance and makes the steps necessary for pre-specifying the analytic design of RWE studies explicit. The *Roadmap* includes steps of defining a study question and the target of estimation, defining the processes that generate data to answer that question, articulating the assumptions required to give results a causal interpretation, selecting appropriate statistical analyses, and pre-specifying sensitivity analyses. Following the

*Roadmap* may lead to either 1) a fully specified analytic study design (including pre-specified analysis plan) that is sufficient to generate high-quality RWE; or, 2) an evidence-based decision that an RWE study to generate the required level of evidence is not currently feasible, with insights into what data would be needed to generate suitable RWE in the future.

The goal of this paper is to disseminate a *Roadmap*-based unifying framework for specifying analytic study designs for RWE generation to an audience of clinical and translational researchers. We provide an overview of the *Roadmap*, including a list of steps to consider when proposing studies that incorporate RWD. Members of the FIORD Working Groups also provide three case studies as companion papers demonstrating application of the *Roadmap* for different use cases, as described in Table 1.

**Table 1: Case Studies Demonstrating Use of the *Roadmap***

| Case Study | Context | Roadmap Steps Emphasized |
|---|---|---|
| Sentinel System and Scalable Phenotyping | Drug safety and monitoring | Outcome-blind[†] simulations to guide estimator pre-specification and machine learning plus natural language processing to enhance identifiability |
| Nifurtimox for Chagas Disease | RCT infeasible | Sensitivity analysis and defining the plausible causal gap |
| Semaglutide and Cardiovascular Outcomes | Secondary indications | Roadmap for hybrid RCT-RWD studies and comparison of complete analytic study designs |

[†]We use outcome-blind to mean without information on the observed treatment-outcome association.

# Overview of the *Causal Roadmap* for clinical and translational scientists

**Figure 1: The *Causal Roadmap***

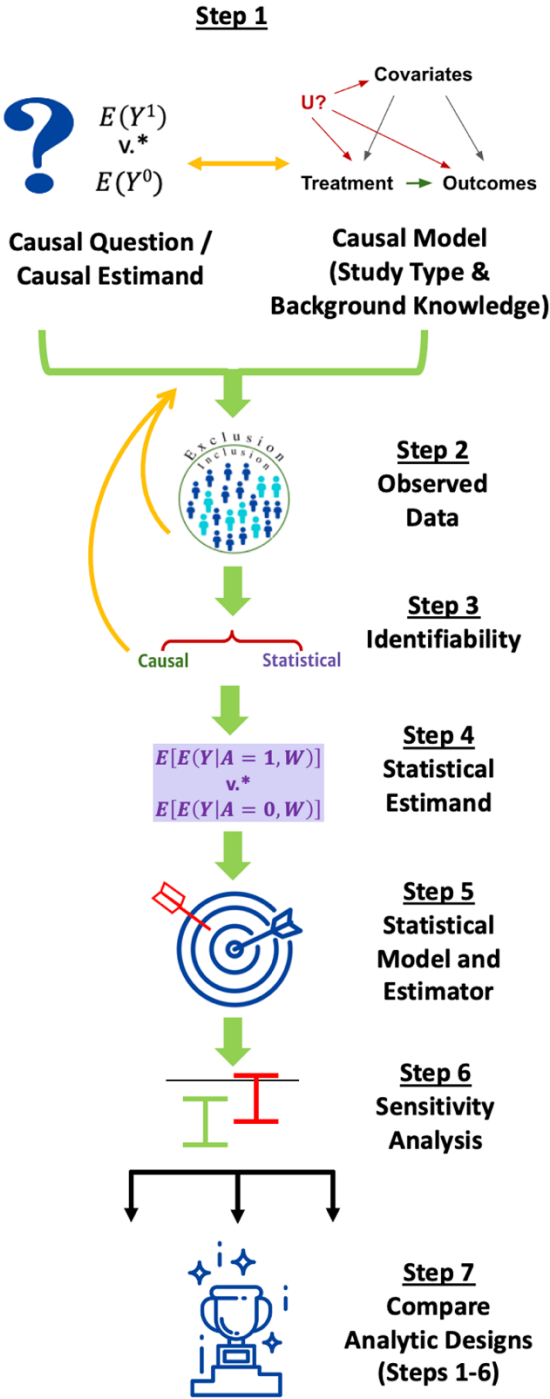

**Caption:** *The contrast of interest may be additive (e.g., risk difference) or multiplicative (e.g., relative risk)

We walk through the steps of the *Roadmap*, depicted in Figure 1, explaining their execution in general terms for simple scenarios, why they are important, and why multidisciplinary collaboration is valuable to accomplish each step. The structured approach outlined in *Roadmap* Steps 1-6 leads to specification of a complete analytic study design, which we define as including not only the type of study (e.g., randomized trial, observational cohort) but also elements of study design from the causal inference literature and the statistical analysis plan. The *Roadmap* does not cover all the steps necessary to write a protocol for running a prospective study, but instead specifies an explicit process for defining the study design itself, including information that is relevant for evaluating the quality of RWE that may be generated by that design. We suggest that following the *Roadmap* can help investigators generate high-quality RWE to answer questions that are important to patients, payers, regulators, and other stakeholders.

A century's worth of literature has contributed to the concepts described in the *Roadmap*. Several books explain nuances of these concepts[24,31–36]. The current paper is not a comprehensive introduction, but rather aims to highlight steps that need to be considered to conduct high-quality causal inference and evidence generation.

**Step 1: Causal question, causal model, and causal estimand**

Step 1 involves defining a causal question, causal model, and the causal estimand that would answer that question. Formally, the causal model that describes relationships between key study variables would be defined before the causal estimand[25]. However, to facilitate explanation of these concepts, we start by using frameworks for specifying components of a causal estimand to also specify key elements of the causal model (Step 1a) before completing our causal model in Step 1b.

**Step 1a: Define the causal question and causal estimand**

Many causal questions start with the objective of estimating the effect of an exposure (e.g., a medication or intervention) on an outcome. Building on decades of research in the careful conduct of randomized and observational studies[36–40], both the International Council for Harmonisation of Technical Requirements for Pharmaceuticals for Human Use (ICH) E9(R1)[41] and Target Trial Emulation[17,32,42,43] frameworks prompt investigators to define components of a causal question and estimand. The causal estimand is a mathematical quantity that answers the causal question (Table 2).

**Table 2: Components of a Causal Question and Estimand per ICH E9(R1)**[41] **and Target Trial Emulation**[17]

| ICH E9(R1) attribute | Target Trial Emulation Protocol Component | Explanation | Related Notation in this Paper |
|---|---|---|---|
| Population | Eligibility criteria | Inclusion and exclusion criteria, including dates of eligibility, for the potential study population | Measured baseline characteristics[†]: $W$ |
| Treatment | Treatment strategies | The ideal hypothetical intervention(s) of interest in each arm of the target trial, including what treatment or exposure or intervention individuals would experience at study baseline <u>and</u> any post-baseline interventions, such as preventing censoring or requiring adherence for a specified duration. | Baseline treatment: $A$, Censoring[††]: $C$ |
| | Follow-up period | The events that define the starting (e.g., randomization, prescription) and stopping (e.g., outcome, death) points for the observation period | |
| Variable or endpoint | Outcome | Outcome of interest, including the timepoint(s) at which the outcome will be evaluated | Outcome: $Y$ |
| Population summary | Causal contrasts of interest | Causal Estimand[†††]: e.g., average treatment effect, causal relative risk, average treatment effect within pre- | See below |

|  |  | specified subgroups |  |

[†]Baseline participant characteristics can include additional variables not used to define eligibility criteria. Baseline variables do not completely characterize the population, but for simplicity, we only consider measured baseline characteristics in the notation below.
[††]In the current paper we focus on interventions on baseline treatment and postbaseline censoring. However, the approach represented extends naturally to treatment strategies that incorporate additional postbaseline interventions, (see e.g., Robins and Hernán (2009)[100], Petersen (2014)[28])
[†††]A mathematical quantity that is a function of potential outcomes (see below).

An example of a question guided by these attributes might be: *How would the risk of disease progression by 2 years have differed if all individuals who met eligibility criteria had experienced treatment strategy A=1 (e.g., drug under investigation) versus treatment strategy A=0 (e.g., active comparator) and no one dropped out of the study (C=0)?* The best (albeit impossible!) way to answer this question would be to evaluate both the potential outcomes[44,45] individuals would have had if they had experienced treatment strategy *A=1* and not been censored ($Y^{a=1,c=0}$) *and* the potential outcomes individuals would have had if they had experienced treatment strategy *A=0* and not been censored ($Y^{a=0,c=0}$).

A formal structural causal model would help us describe the causal pathways that generate these potential outcomes[46]. For now, we simply consider that, if we were able to observe both potential outcomes for all members of our target population, then the answer to our question would be given by the causal risk difference (or "Average Treatment Effect"),

$$\Psi^* = P(Y^{a=1,c=0} = 1) - P(Y^{a=0,c=0} = 1).$$

This mathematical quantity that is a function of potential outcomes is called a *causal estimand*. Table 2 lists other examples of causal estimands.

**Importance:** Even though we can only observe at most one potential outcome for each individual[47], and even though it is not possible to guarantee complete follow-up in a real trial, precise definition of the causal question and estimand based on the treatment strategies defined in Table 2 is crucial for specifying a study design and analysis plan to provide the best possible effect estimate. Ultimately, we need to evaluate a mathematical expression that translates the available data into a number (e.g., a 5% decrease in risk of disease progression). To assess whether that number provides an answer to our causal question, we must first define mathematically what we aim to estimate.

**Build a Multidisciplinary Collaboration:** If you are not certain how to translate your research question into a causal estimand, collaborate with an expert in causal inference.

**Step 1b: Specify a causal model describing how data have been or will be generated**

Next, we consider what we know (and don't know) about the real-life processes that will generate – or that have already generated – data to answer this question. First, we consider the type of study (e.g., pragmatic RCT, retrospective cohort study). Then, we consider what factors affect the variables that are part of our treatment strategies – found in Table 2 and referred to as intervention variables below – and the outcome in our proposed study.

This background knowledge comprises the causal model[46]. We specified some key variables in our causal model in Step 1a (in Table 2 and our potential outcomes). Now, we add additional detail to our causal model by describing potential causal relationships between these and other important variables. Multiple tools and frameworks can help elicit this information, but conceptual models and causal graphs, such as directed acyclic graphs or single world intervention graphs, are some of the most common[39,48–51].

Figure 2 gives a simple example of causal graph construction, starting with writing down all intervention and outcome variables. When some outcomes are missing, we don't observe the outcome, $Y$, for all participants. Instead, we observe $Y^*$, which is equal to the actual outcome if it was observed and is missing otherwise (Figure 2a). Arrows denote possible effects of one variable on another.

**Figure 2: Basic Process for Generating a Causal Graph**

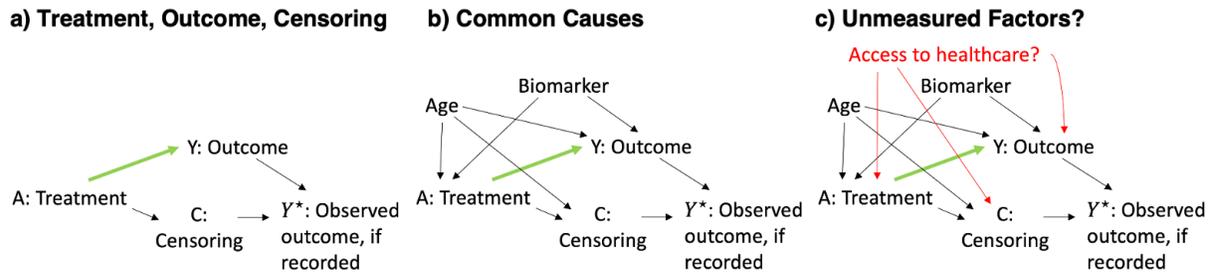

**Caption:** $Y^*$ is equal to the actual outcome value if it was observed and is missing otherwise.

Then, we attempt to write down factors that might influence these variables. Figure 2b shows two examples (age and a biomarker), though real causal graphs generally include many more variables. In a classic randomized trial, only the randomization procedure affects baseline treatment assignment, whereas in an observational study (depicted in Figure 2), participant characteristics affect the baseline treatment. Next, we consider factors that are unmeasured or difficult to measure that might influence treatment, outcomes, or censoring. Figure 2c shows access to healthcare as an example.

Causal graphs can become much more complicated, especially when working with longitudinal data[52], using proxies for unmeasured variables[53], or combining different data sources[54] (as demonstrated in the case study of Semaglutide and Cardiovascular Outcomes). A carefully constructed causal graph should also demonstrate issues such as competing risks, intercurrent events, or measurement error[32,55].

**Importance:** Considering which factors may affect intervention variables and outcomes helps to determine whether we can answer our question based on existing data or data that we will collect. The final graph should be our best honest judgement based on available evidence and incorporating remaining uncertainty[32].

**Build a Multidisciplinary Collaboration:** If questions remain about some aspect of this model, such as how physicians decide to prescribe a medication in different practice settings, obtain input from clinicians or other relevant collaborators before moving on.

**Stop! Do you need to modify your causal question and estimand (Step 1a) based on Step 1b?**

After writing down our causal model, we sometimes need to change our question[56]. For example, we may have realized that an intercurrent event (such as death) prevents us from observing the outcome for some individuals. As suggested by ICH E9(R1), we could modify the question to consider the effect on a composite outcome of the original Y or death[41]. ICH E9(R1)[41] discusses other intercurrent events and potential modifications to the estimand.

**Step 2: Describe the observed data**

The causal model from Step 1b lets us specify what we know about the real-world processes that generate our observed data. This model can inform what data we collect in a prospective study or help to determine whether existing data sources include relevant information. Next, we describe the actual data we will observe.

Specific questions to address regarding the observed data include the following: How are the relevant exposures, outcomes, and covariates, including those defining eligibility criteria, measured in the observed data? Are they measured differently (including different monitoring protocols) in different data sources or at different timepoints? Are we able to measure all variables that are important common causes of the intervention variables and the outcome? Is the definition of time zero in the data consistent with the causal question[42]?

**Importance:** After considering these questions, we may need to modify Step 1. For example, if we realize that the data we are able to observe only include patients seen at tertiary care facilities, we may need to change the question (Step 1a) to ask about the difference in the risk of disease progression by two years if all individuals meeting our eligibility criteria *and receiving care at a tertiary facility* received one intervention or the other. Knowledge about factors that affect how variables are measured and whether they are missing should be incorporated in the causal model (Step 1b). Completing this step also helps investigators assess whether the data are fit-for-use[3] and whether we are able to estimate a causal effect from the observed data (discussed in Step 3).

**Build a Multidisciplinary Collaboration:** If you are unsure about the way variables are measured in relation to underlying medical concepts or in relation to a particular care setting, collaborate with a clinician or clinical informaticist. If you are unsure of how to match baseline time zero in your observed data with the follow-up period in your causal question, collaborate with a statistician.

**Step 3: Assess identifiability: Can the proposed study provide an answer to our causal question?**

In Step 3, we ask whether the data we do observe (Step 2), together with our knowledge about how these data are generated (Step 1b), are sufficient to let us answer our causal question (Step 1a). As described in Step 1a, we cannot directly estimate our causal estimand (which is a function of counterfactual outcomes). Instead, we will evaluate a function of the observed data (called a statistical estimand, described in Step 4). The difference between the true values of the statistical and causal estimands is sometimes referred to as the causal gap[27]. If there is a causal gap, even a perfect estimate of the statistical estimand would not provide an answer to our causal question.

While we can never be certain of the size of the causal gap for studies incorporating RWD and even for many questions using data from traditional RCTs, we must use our background knowledge to provide an honest appraisal. Causal identification assumptions help us to explicitly state what must be true in order to conclude that the causal gap is zero and that we are thus able to estimate a causal effect using the proposed data. Table 3 lists two common identification assumptions to consider for most cases with informal explanations of their meaning. Exchangeability, in particular, can also be framed in terms of causal graphs[46]. In some cases, further assumptions may be necessary. Hernán and Robins (2020)[32], among others, provide in-depth discussions of identification assumptions. The case studies associated with this paper demonstrate the evaluation of these assumptions.

**Table 3: Common Identification Assumptions**

| Assumption | Basic Explanation of Meaning |
|---|---|
| Exchangeability[†] | This assumption is generally true if there are no unmeasured common causes of variables that are part of the treatment strategies (Table 2: e.g., baseline or postbaseline treatment(s), censoring) and the outcome (informally, if there is no unmeasured confounding). |
| Positivity | This assumption is true if, for every possible combination of measured confounding variables, individuals with those characteristics have a positive probability of following any of the treatment strategies of interest. |

[†]Full exchangeability is generally not required if weaker conditions (e.g., mean exchangeability, sequential conditional exchangeability, or others) hold[32].

**Importance:** Considering and documenting the plausibility of the causal identification assumptions helps to determine whether steps can be taken to decrease the potential magnitude of the causal gap. If we conclude that these assumptions are unlikely to be satisfied, then we should consider modifications to Steps 1-2. We may need to limit the target population to those who have a chance of receiving the intervention or evaluate the effect of a more realistic treatment rule to improve the plausibility of the positivity assumption[57,58]. We may need to measure more of the common causes depicted in our causal graph or modify the question to improve the plausibility of the exchangeability assumption[59]. If multiple study designs are feasible, Step 3 can help us to consider which study design is based on more reasonable assumptions[60].

If we know that a key variable affecting treatment and outcomes or censoring and outcomes is not measured, then we generally cannot identify a causal effect from the observed data without measuring that variable or making additional assumptions[17,32,37]. For this and other reasons, many studies analyzing RWD appropriately report statistical associations and not causal effects, though sensitivity analyses (Step 6) may still help to evaluate whether a causal effect exists[61,62]. Nonetheless, if a retrospective study was initially proposed but the causal identification assumptions are highly implausible and cannot be improved using existing data, then investigators should consider prospective data collection to better evaluate the effect of interest.

In general, it would be unreasonable to expect that all causal identification assumptions would be exactly true in RWE studies, or even in many traditional RCTs due to issues such as informative missingness[32]. Yet careful documentation of Steps 1-3 in the pre-specified analysis plan and in the study report helps not only the investigator but also regulators, clinicians, and other stakeholders to evaluate the quality of evidence generated by the study about the causal effect of interest. Step 3 helps us to specify a study with the smallest causal gap possible. Sensitivity analyses, discussed in Step 6, help to quantify a reasonable range for the causal gap, further aiding in the interpretation of RWE study results.

**Build a Multidisciplinary Collaboration:** An expert in causal inference can help to formally evaluate all causal identification assumptions. The exchangeability assumption can become quite complicated if there are multiple intervention variables[39,52]. In such cases, graphical criteria may be used to determine visually from a causal graph whether sufficient variables have been measured to satisfy the exchangeability assumption[39,46,63]. Software programs can also facilitate this process[64,65].

### Step 4: Define the statistical estimand

If, after assessing identifiability, we decide to proceed with our study, we aim to define a statistical estimand that is as close as possible to the causal estimand of interest. Recall our causal risk difference for a single time-point intervention and outcome:

$$\Psi^* = P(Y^{a=1,c=0} = 1) - P(Y^{a=0,c=0} = 1).$$

In a simple case where participant characteristics other than our intervention variables and outcome – denoted $W$ – are only measured at baseline, then the statistical estimand that is equivalent to the causal effect if all identification assumptions are true is given by

$$\Psi = E_W(P[Y^\star | C = 0, A = 1, W] - P[Y^\star | C = 0, A = 0, W]).$$

In words, we have re-written our causal question (which is defined based on potential outcomes that we cannot simultaneously observe) in terms of a quantity that we can estimate with our data: the average (for our target population) of the difference in risk of our observed outcome associated with the different treatment strategies, adjusted for measured confounders.

**Importance:** The traditional practice of defining the statistical estimand as a coefficient in a regression model has several downsides, even if the model is correct (a questionable assumption discussed below)[24]. This approach starts with a tool (e.g., a regression model) and then asks what problem it can solve, rather than starting with a problem and choosing the best tool[66]. For example, the hazard ratio may be estimated based on a coefficient in a Cox regression but does not correspond to a clearly defined causal effect[67–69]. Instead, the *Roadmap* guides us to choose a statistical estimand that is as close as possible to the causal estimand. We thus specify a well-defined quantity that can be estimated from the observed data and that is directly linked to the causal question.

**Build a Multidisciplinary Collaboration:** Defining a statistical estimand that would be equivalent to the causal effect of interest under identification assumptions is more challenging when there are post-baseline variables that are affected by the exposure and that, in turn, affect both the outcome and subsequent intervention variables[39]. This situation is common in studies where the exposure is measured at multiple time-points. In such a situation, statistician collaborators can help to define the statistical estimand using approaches such as the longitudinal g-computation formula[39].

**Step 5: Choose a statistical model and estimator that respects available knowledge and uncertainty based on statistical properties**

The next step is to define a statistical model (formally, the set of possible data distributions) and to choose a statistical estimator. The statistical model should be compatible with the causal model (Step 1b). For example, knowledge that treatment will be randomized (design knowledge that we described in our causal model) implies balance in baseline characteristics across the two arms (with slight differences due to chance in a specific study sample). We could also incorporate knowledge that a continuous outcome falls within a known range or that a dose-response curve is monotonic (e.g., based on prior biological data) into our statistical model. A good statistical model summarizes such statistical knowledge about the form of the relationships between observed variables that is supported by available evidence without adding any unsubstantiated assumptions (such as linearity, or absence of interactions); models of this type are often referred to as semi- or non-parametric or simply realistic statistical models[24].

Given a statistical model, the choice of estimator should be based on pre-specified statistical performance benchmarks that evaluate how well it is likely to perform in estimating the statistical estimand[24]. Examples include type 1 error control, 95% confidence interval (CI) coverage, statistical bias, and precision. Statistical bias refers to how far the average estimate across many samples would be from the true value of the statistical estimand. An estimator must be flexible enough to perform well even when we do not know the form of the association between variables in our dataset, and it must be fully pre-specified[24].

Most available estimators rely on estimating an outcome regression (i.e., the expected value of the outcome given the treatment and values of confounders), a propensity score (i.e., the probability of receiving a treatment or intervention given the measured confounders), or both. Without knowing the form of these functions, we do not know *a priori* whether they are more likely to be accurately modeled with a parametric regression or a flexible machine learning algorithm allowing for non-linearities and interactions between variables[24,66,70]. The traditional practice of defaulting to a parametric regression as the statistical estimator imposes additional untestable statistical assumptions, even though they are not necessary. Fortunately, estimators exist that allow for full pre-specification of all machine learning and parametric approaches used, data-adaptive selection (e.g., cross-validation) of the algorithm(s) that perform best for a given dataset, and theoretically-sound 95% confidence interval construction (leading to proper coverage under reasonable conditions)[24].

**Importance:** Effect estimates that are based on incorrectly specified models – such as a main terms linear regression when there is truly non-linearity or interactions between variables – are biased, and that bias does not get smaller as sample size increases[24].

This bias may result in inaccurate conclusions. We aim to choose an estimator that not only has minimal bias but also is efficient – thereby producing 95% confidence intervals that are accurate but as narrow as possible – to make maximal use of the data[24].

If, after consideration of the statistical assumptions and properties of the estimators, multiple estimators are considered, then the bias, variance, and 95% CI coverage of all estimators should be compared using outcome-blind simulations that mimic the true proposed experiment as closely as possible[71]. We use outcome-blind to mean that the simulations are conducted without information on the observed treatment-outcome association; such simulations may utilize other information from the collected data (if available), such as data on baseline covariates, treatment, and censoring, to approximate the real experiment[71]. Simulations conducted before data collection may use a range of plausible values for these study characteristics[72]. As recommended by ICH E9(R1), simulations should also be conducted for cases involving plausible violations of the statistical assumptions of the estimators[41]. Examples of such violations include non-linearity for linear models or inaccurate prior distributions for Bayesian parameters. For an example of conducting such a simulation, please see the Drug Safety and Monitoring case study.

**Build a Multidisciplinary Collaboration:** Statistician collaborators can help to pre-specify an estimator with the statistical properties described above. Resources are increasingly available to assist with pre-specification of statistical analysis plans (SAPs) based on state-of-the-art estimation approaches. For example, Gruber et al. (2022)[73] provide a detailed description of how to pre-specify a SAP using targeted minimum loss-based estimation (TMLE)[74] and super learning[70], a combined approach that integrates machine learning to minimize the chance that statistical modeling assumptions are violated[24].

**Step 6: Specify a procedure for sensitivity analysis**

Sensitivity analyses in Step 6 attempt to quantify how the estimated results (Step 5) would change if the untestable causal identification assumptions from Step 3 were violated[32,61,75–77]. In contrast, the simulations in Step 5 consider bias due to violations of testable statistical assumptions, which ICH E9(R1) considers as a different form of sensitivity analysis[41]. One mechanism of conducting a causal sensitivity analysis in Step 6 is to consider the potential magnitude and direction of the causal gap; this process requires subject matter expertise and review of prior evidence[61,76–78]. Sensitivity analysis also allows for construction of confidence intervals that account for plausible values of the causal gap[27,61,76–78]. Alternatively, investigators may assess for causal bias using negative control variables, discussed in detail by Lipsitch et al. (2010)[79] and Shi et al. (2020)[80].

The specifics of these methods – and alternate approaches – are beyond the scope of this paper, but the case study of Nifurtimox for Chagas Disease provides an overview of methods for sensitivity analysis, as well as a worked example of using available evidence to assess a plausible range for the causal gap. As discussed in this case study, the method of sensitivity analysis should be pre-specified prior to estimating the effect of interest[81]. This process avoids the bias that might occur if experts know the value of the estimate before defining the procedure they will use to decide whether a given shift in that estimate due to bias is reasonable[76].

**Importance:** The process of using prior evidence to reason about likely values of the causal gap helps investigators to assess the plausibility that the bias due to a violation of identification assumptions could be large enough that the observed effect is negated[27,61,62,82]. While the exact magnitude of the causal effect may still not be identified due to known issues such as the potential for residual confounding, if an estimated effect is large enough, we may still obtain credible evidence that an effect exists[62,83]; this was the case in Cornfield et al. (1959)'s frequently-cited assessment of the effect of smoking on lung cancer[84]. Conversely, if the anticipated effect size is small and the plausible range of the causal gap is large, the proposed study may not be able to provide actionable information. Considering these tradeoffs can help investigators to decide whether to pursue a given RWE study or to consider alternate designs that are more likely to provide high-quality evidence of whether a causal effect exists[62,85].

**Build a Multidisciplinary Collaboration:** If multiple correlated sources of bias are likely, more complex methods of evaluating a plausible range for the causal gap – and collaboration with investigators familiar with these methods – may be required[76].

**Step 7: Compare alternative complete analytic study designs**

*Roadmap* Steps 1-6 help us to specify a complete analytic study design, including the causal question and estimand, type of study and additional knowledge about how the data are generated, specifics of the data sources that will be collected and/or analyzed, assumptions that the study relies on to evaluate a causal effect, statistical estimand, statistical estimator, and procedure for sensitivity analysis. The type of study described by this analytic design could fall anywhere on the spectrum from a traditional RCT to a fully observational analysis. In cases when it is not possible to conduct a traditional RCT due to logistical or ethical reasons – or when RCT results would not be available in time to provide actionable information – the value of RWE studies is clear despite the possibility of a causal gap[32]. If an RCT is feasible, baseline randomization of an intervention (as part of either a traditional or pragmatic RCT[86]) still generally affords a higher degree of certainty that the estimated effect is causal compared to analysis of non-randomized data. Yet sometimes, it is feasible to consider multiple different observational and/or randomized designs – each with different potential benefits and downsides.

Consider a situation in which there is some evidence for a favorable risk-benefit profile of a previously studied intervention based on prior data, but those data are by themselves insufficient for regulatory approval for a secondary indication or for clear modification of treatment guidelines. In this context, it is possible that conducting a well-designed RWE study or hybrid RCT-RWD study as opposed to a traditional RCT alone will shorten the time to a definitive conclusion, decrease the time patients are exposed to an inferior product, or provide other quantifiable benefits to patients while still providing acceptable control of type I and II errors[87–89]. Yet other times, a proposed RWE design may be inferior to alternative options, or one design may not be clearly superior to another. When multiple study designs are considered, outcome-blind simulations consistent with our description of Steps 1-6 can help to compare not only type 1 error and power, but also metrics quantifying how the proposed designs will modify the medical product development process[87]. The case study of Semaglutide and Cardiovascular Outcomes demonstrates how to compare study designs that are based on *Roadmap* Steps 1-6.

**Importance:** A simulated comparison is not always necessary; one study design may be clearly superior to another. Yet often there are tradeoffs between studies with different specifications of *Roadmap* Steps 1-6. For example, in some contexts, we may consider augmenting an RCT with external data. When comparing the RCT design to the augmented RCT design, there may be a tradeoff between a) the probability of correctly stopping the study early when appropriate external controls are available and b) the worst-case type 1 error that would be expected if inappropriate external controls are considered[89]. Another example would be the tradeoff between the potential magnitudes of the causal gap when different assumptions are violated to varying degrees for studies relying on alternate sets of causal identification assumptions[60]. Simulated quantification of these tradeoffs using pre-specified benchmarks can help investigators to make design choices transparent[90].

**Build a Multidisciplinary Collaboration:** Factors to consider when comparing different analytic designs include the expected magnitude of benefit based on prior data and the quality of that data[11], the plausible bounds on the causal gap for a given RWE study, the treatments that are currently available[11], and preferences regarding tradeoffs between design characteristics such as type I versus type II error control[90]. Because these tradeoffs will be context-dependent[11,90], collaboration with patient groups and discussion with regulatory agencies is often valuable when choosing a study design from multiple potential options.

**A list of *Roadmap* steps for specifying a complete analytic study design**

Table 4 provides a list of considerations to assist investigators in completing and documenting all steps of the *Roadmap*. Complete reporting of RWE study results should include all pre-specified *Roadmap* steps, though information supporting decisions in the final design and analysis plan, such as causal graphs or simulations, may be included as supplementary material. Note that all steps should be pre-specified before conducting the study.

**Table 4: Steps for Specifying a Complete Analytic Study Design Using the *Roadmap***

| | Roadmap Step |
|---|---|
| 1a | ● Specify the causal question and estimand.<br>　● ICH E9(R1) attributes: Population, treatment, variable or endpoint, population summary[41]<br>　● Target Trial Emulation Protocol Components: Eligibility criteria, treatment strategies, follow-up period, outcome, causal contrasts of interest [32] |
| 1b | ● Specify the causal model (based on background knowledge about the proposed study).<br>　● Specify the type of study (e.g., traditional RCT, retrospective cohort)<br>　● Document whether censoring, competing risks, or other intercurrent events occurred and factors that may have affected them. Adjust the question as needed. |
| 2 | ● Define the observed data that will be or has been collected.<br>　● Document how the inclusion/exclusion criteria, treatment variables, outcome(s), and other relevant variables are measured, how time zero is defined, and important differences between data sources. |
| 3 | ● Assess identifiability of the causal estimand from the observed data.<br>　● Explicitly state the assumptions required for identification, and evaluate their plausibility.<br>　● Consider modifications to Steps 1-2 to minimize the causal gap.<br>　　● If a retrospective study had been planned but identification assumptions are highly implausible, consider primary data collection or linkage of data from different sources as necessary to ensure relevant information capture for the causal question and estimand. |
| 4 | ● Define the statistical estimand. |
| 5 | ● Specify the statistical model, estimator, and method of confidence interval construction.<br>　● List the assumptions the proposed estimator and method of confidence interval construction rely upon.<br>　● Describe the expected statistical bias and variance of the estimator under plausible conditions.<br>　● If multiple estimators are considered, compare them with outcome-blind simulations based on:<br>　　● statistical bias, variance, 95% CI coverage of the statistical estimand, type 1 error, and power |
| | • with plausible violations of model assumptions. |
| 6 | • Specify the sensitivity analyses.<br>  • Document the method for defining plausible bounds for the causal gap and/or methods for estimation of the causal gap (e.g., based on negative controls).<br>  • Provide confidence intervals for the causal effect of interest under the hypothesized size of the causal gap, across the full range of plausible causal gaps. |
| 7 | • Compare feasible complete analytic designs (Steps 1-6) using outcome-blind simulations based on:<br>  • causal metrics (95% CI coverage, type 1 error, and power for a causal effect),<br>  • and metrics to quantify differences in the medical product development process of each design.<br>  • Include a comparison to an RCT if feasible. |

**Discussion**

The *Roadmap* can help investigators to pre-specify a complete analytic design for studies that utilize RWD, choose between study designs, and propose high-quality RWE studies to the FDA and other agencies. We describe the steps of the *Roadmap* in order to disseminate this methodology to clinical and translational scientists. The case study companion papers on Drug Safety and Monitoring, Nifurtimox for Chagas Disease, and Semaglutide and Cardiovascular Outcomes demonstrate applications of the *Roadmap*, with each study explaining specific steps in greater detail.

Past descriptions of the *Roadmap* have largely been targeted to quantitative scientists[24–27,29,30]. In this manuscript, we focus on intuitive explanations rather than formal mathematical rigor to make these causal inference concepts more accessible to a wide audience. We also emphasize the importance of building a multidisciplinary collaboration, including both clinicians and statisticians, during the study planning phase.

We also introduce an extension of previous versions of the *Roadmap* to emphasize how outcome-blind simulations may be used not only to compare different statistical estimators but also to evaluate different study designs. This extension aligns with the FDA's Complex Innovative Trial Designs Program guidance for designs that require simulation to estimate type I and II error rates[91], but goes one step further by emphasizing a quantitative comparison to a randomized trial or other feasible RWE designs. The aim of this additional step is to facilitate evaluation of the strengths and weaknesses of each potential approach.

The *Roadmap* aligns with other recommendations provided in regulatory guidance, as well; these include the FDA's Framework and Draft Guidance documents for RWE that emphasize the quality and appropriateness of the data[3,92–94] and the ICH E9(R1) guidance on estimands and sensitivity analysis[41]. The *Roadmap* is also consistent with other proposed frameworks for RWE generation. Within the field of causal inference, the *Roadmap* brings together concepts including potential outcomes[44,45], the careful design of non-experimental studies[35,36,38,40], causal graphs[39,48–51] and structural causal models[46], causal identification[39,46,95], translation of causal to statistical estimands using the g-formula[39], and methods for estimation and sensitivity analysis[24,34,61,70,75,77].

The *Roadmap* is compatible with many other frameworks, including many that discuss aspects of specific *Roadmap* steps. Examples include the Target Trial Emulation framework[17,43], the Patient-Centered Outcomes Research Institute (PCORI) Methodology Standards[19], white papers from the Duke-Margolis Center[18,96], the REporting of studies Conducted using Observational Routinely-Collected health Data (RECORD) Statement[97], the Structured Preapproval and Postapproval Comparative study design framework[98], and the STaRT-RWE template[20]. The purpose of the *Roadmap* is not to replace these – and many other – useful sources of guidance, but rather to provide a unified framework that covers the steps necessary to follow a wide range of guidance in a centralized location. Furthermore, while many recommendations for RWE studies list *what* to think about (e.g., types of biases or considerations for making RWD and trial controls comparable), the *Roadmap* aims instead to make explicit a process for *how* to make and report design and analysis decisions that is flexible enough to be applied to any use case along the spectrum from a traditional RCT to a fully observational analysis.

With increasing emphasis by regulatory agencies around the world regarding the importance of RWE[5], the number of studies using RWD that contribute to regulatory decisions is likely to grow over time. Yet a recent review of RWE studies reported that "nearly all [reviewed] studies (95%) had at least one avoidable methodological issue known to incur bias"[99]. By following the *Roadmap* steps to fully pre-specify an analytic study design, investigators may set themselves up not only to convey relevant information to regulators but also to produce high-quality estimates of causal effects using RWD when possible, with an honest evaluation of whether the proposed study is adequate for making causal inferences.

**Disclaimer:** The contents are those of the author(s) and do not necessarily represent the official views of, nor an endorsement by, FDA/HHS, the U.S. Government, or the authors' affiliations.

**Acknowledgements:** We would like to thank the sponsors of the FIORD workshop, including the Forum for Collaborative Research and the Center for Targeted Machine Learning and Causal Inference (both at the School of Public Health at the University of California, Berkeley), and the Joint Initiative for Causal Inference. We would also like to thank Dr. John Concato for his comments on this manuscript. This research was funded by a philanthropic gift from the Novo Nordisk corporation to the University of California, Berkeley to support the Joint Initiative for Causal Inference.

**Disclosures: LED** reports tuition and stipend support from a philanthropic gift from the Novo Nordisk corporation to the University of California, Berkeley to support the Joint Initiative for Causal Inference. **EK** is employed by Microsoft. **DA** is employed by Pfizer Inc. and holds stocks in Pfizer Inc. **CYV** and **RAH** are employed by Syneos Health. **CYV** reports that her husband is employed by Galapagos. **HR** and **KK** are employed by Novo

Nordisk A/S and own stocks in Novo Nordisk A/S. **RP** has received the following (directed to his institution): speaker fees from Merck and Novo Nordisk; consulting fees from Bayer AG, Corcept Therapeutics Incorporated, Dexcom, Endogenex, Inc., Gasherbrum Bio, Inc., Hanmi Pharmaceutical Co., Hengrui (USA) Ltd., Lilly, Merck, Novo Nordisk, Pfizer, Rivus Pharmaceuticals Inc., Sanofi, Scohia Pharma Inc., and Sun Pharmaceutical Industries; and grants from Hanmi Pharmaceuticals Co., Metavention, Novo Nordisk, and Poxel SA. **DM** is employed by Moderna. **CEB** is co-founder of Adaptic Health Inc., Managing Director of Pivotal Strategic Consulting, LLC, and receives consulting fees from Graticule Inc. and Sophic Alliance Inc. **MA** is employed by Novartis Pharma AG. **JBB** reports contracted fees and travel support for contracted activities for consulting work paid to the University of North Carolina by Novo Nordisk; grant support by Dexcom, NovaTarg, Novo Nordisk, Sanofi, Tolerion and vTv Therapeutics; personal compensation for consultation from Alkahest, Altimmune, Anji, AstraZeneca, Bayer, Biomea Fusion Inc, Boehringer-Ingelheim, CeQur, Cirius Therapeutics Inc, Corcept Therapeutics, Eli Lilly, Fortress Biotech, GentiBio, Glycadia, Glyscend, Janssen, MannKind, Mellitus Health, Moderna, Pendulum Therapeutics, Praetego, Sanofi, Stability Health, Terns Inc, Valo and Zealand Pharma; stock/options in Glyscend, Mellitus Health, Pendulum Therapeutics, PhaseBio, Praetego, and Stability Health; and board membership of the Association of Clinical and Translational Science. **MvdL and SG** report that they are co-founders of the statistical software start-up company TLrevolution, Inc. **MvdL and MP** report personal compensation for consultation from Novo Nordisk.